\documentclass[a4paper,fleqn,usenatbib]{mnras}

\usepackage{color}

\usepackage[T1]{fontenc}
\usepackage{ae,aecompl}

\usepackage{graphicx}	
\usepackage{amsmath}	
\usepackage{amssymb}	
\usepackage{multirow}
\usepackage{soul}
\usepackage{float}

\title[AGN-enhanced outflows in SFGs at 1.7<z<4.6]{AGN-enhanced outflows of low-ionization gas in star-forming galaxies at 1.7<z<4.6
	  \thanks{Based on data obtained with the European Southern Observatory Very Large Telescope, Paranal, Chile, under Programs 170.A-0788, 074.A-0709 and 275.A-5060 (ESO-GOODS/FORS2), 164.O-0560 (K20), 072.A-0139, 080.A-0411, 66.A-0270 and 67.A-0418 (spectroscopic follow-up of X-ray sources in the Chandra Deep Field South and Extended Chandra Deep Field South), 173.A-0687 (GMASS), 171.A-3045 (ESO-GOODS/VIMOS), 70.A-9007 (VVDS), 175.A-0839 (zCOSMOS), and 185.A-0791 (VUDS); and based on data collected at the Magellan Telescope, which is operated by the Carnegie Observatories.}
}

\author[M. Talia et al.]{M.~Talia,$^{1,2}$\thanks{E-mail: margherita.talia2@unibo.it}
M.~Brusa,$^{1,2}$
A.~Cimatti,$^{1,15}$
B. C.~Lemaux,$^{3,4}$
R.~Amorin,$^{5,13,14}$
S.~Bardelli,$^{2}$
\newauthor
L. P.~Cassar\`a,$^{6}$
O.~Cucciati,$^{2}$
B.~Garilli,$^{7}$
A.~Grazian,$^{5}$
L.~Guaita,$^{5}$
N. P.~Hathi,$^{8,4}$
\newauthor
A.~Koekemoer,$^{8}$
O.~Le F\`evre,$^{4}$
D.~Maccagni,$^{7}$
K.~Nakajima,$^{9}$
L.~Pentericci,$^{5}$
J.~Pforr,$^{10,4}$
\newauthor
D.~Schaerer,$^{9,11}$
E.~Vanzella,$^{2}$
D.~Vergani,$^{2,12}$
G.~Zamorani,$^{2}$
E.~Zucca$^{2}$
\\
$^{1}$Dipartimento di Fisica e Astronomia, Universit\`a di Bologna, Via Gobetti 93/2, I-40129, Bologna, Italy\\
$^{2}$INAF- Osservatorio Astronomico di Bologna, Via Gobetti 93/3, I-40129, Bologna, Italy\\
$^{3}$Department of Physics, University of California, Davis, One Shields Ave., Davis, CA 95616, USA\\
$^{4}$Aix Marseille Universit\'e, CNRS, LAM (Laboratoire d'Astrophysique de Marseille) UMR 7326, 13388, Marseille, France\\
$^{5}$INAF--Osservatorio Astronomico di Roma, via di Frascati 33, I-00040, Monte Porzio Catone, Italy\\
$^{6}$Institute for Astronomy, Astrophysics, Space Applications and Remote Sensing, National Observatory of Athens, Penteli,15236 Athens, Greece\\
$^{7}$INAF--IASF Milano, via Bassini 15, I--20133, Milano, Italy\\
$^{8}$Space Telescope Science Institute, 3700 San Martin Drive, Baltimore MD 21218, USA\\
$^{9}$Geneva Observatory, University of Geneva, ch. des Maillettes 51, CH-1290 Versoix, Switzerland\\
$^{10}$Scientific Support Office, Directorate of Science and Robotic Exploration, European Space Research and Technology Centre (ESA/ESTEC), \\Keplerlaan 1, 2201 AZ Noordwijk, The Netherlands\\
$^{11}$Institut de Recherche en Astrophysique et Plan\'etologie - IRAP, CNRS, Universit\'e de Toulouse, UPS-OMP, 14, avenue E. Belin, \\F31400 Toulouse, France\\
$^{12}$INAF--IASF Bologna, via Gobetti 101, I--40129,  Bologna, Italy\\
$^{13}$Cavendish Laboratory, University of Cambridge, 19 JJ Thomson Avenue, Cambridge, CB3 0HE, UK\\
$^{14}$Kavli Institute for Cosmology, University of Cambridge, Madingley Road, Cambridge CB3 0HA, UK \\
$^{15}$INAF - Osservatorio Astrofisico di Arcetri, Largo E. Fermi 5, I-50125, Firenze, Italy \\
}

\date{Accepted XXX. Received YYY; in original form ZZZ}

\pubyear{2017}

\begin{document}
\label{firstpage}
\pagerange{\pageref{firstpage}--\pageref{lastpage}}
\maketitle

\begin{abstract}
   {Fast and energetic winds are invoked by galaxy formation models as essential processes in the evolution of galaxies. These outflows can be powered either by star-formation and/or AGN activity, but the relative dominance of the two mechanisms is still under debate. We use spectroscopic stacking analysis to study the properties of the low-ionization phase of the outflow in a sample of 1330 star-forming galaxies (SFGs) and 79 X-ray detected ($10^{42}{<}L_{X}{<}10^{45}$erg s$^{-1}$) Type 2 AGN at $1.7{<}z{<}4.6$ selected from a compilation of deep optical spectroscopic surveys, mostly zCOSMOS-Deep and VUDS. 

We measure mean velocity offsets of ${\sim}-150$~km s$^{-1}$ in the SFGs while in the AGN sample the velocity is much higher (${\sim}-950$~km s$^{-1}$), suggesting that the AGN is boosting the outflow up to velocities that could not be reached only with the star-formation contribution. The sample of X-ray AGN has on average a lower SFR than non-AGN SFGs of similar mass: this, combined with the enhanced outflow velocity in AGN hosts, is consistent with AGN feedback in action. 

We further divide our sample of AGN into two X-ray luminosity bins: we measure the same velocity offsets in both stacked spectra, at odds with results reported for the highly ionized phase in local AGN, suggesting that the two phases of the outflow may be mixed only up to relatively low velocities, while the highest velocities can be reached only by the highly ionized phase.}
\end{abstract}




\section{Introduction}\label{sec:Introduction}
Fast and energetic winds are invoked by many galaxy formation models as essential processes in the formation and evolution of galaxies. 
Their existence is well established observationally, though many questions remain to be answered about their origin and nature. One of the most important answers that is sought regards the main engine driving galactic-scale winds, especially at the peak of both SF and AGN activity through cosmic time ($1{<}z{<}3$), in an attempt to explain the rapid quenching of SF and the link between the evolution of AGN and their host galaxies through the feedback that winds could in principle provide.

Indeed, massive gas outflows can be powered either by star-formation (SF) activity, namely stellar winds and/or supernovae explosions, or by AGN, for example through a radiatively-driven process likely associated with a luminous, obscured, and dust-enshrouded black hole accretion phase, though the relative dominance and efficiency of the different mechanisms is not yet fully understood \citep{chevalier1985, king2005, veilleux2005, fabian2012}.

Winds are observed in SFGs with no sign of AGN activity in their UV or optical spectra \citep[e.g.][]{shapley2003, talia2012, cicone2016, bordoloi2013} or from their X-ray properties \citep{cimatti2013}, in quiescent galaxies hosting low-luminosity AGN \citep{cheung2016} and in different types of AGN/QSO \citep[e.g.][]{hainline2011, cimatti2013, cicone2014, brusa2015a, harrison2012, forsterschreiber2014, canodiaz2012, rupke2013b, feruglio2010} spanning a wide range of velocities. 
The comparison of different studies of the outflow phenomenon at high redshift (z$>$1) in normal SFGs (i.e. not starburst) and AGN hosts shows that AGN are generally characterized by faster winds and this velocity difference with respect to SFGs can be interpreted as evidence that the nuclear activity is influencing the gas motion on galactic scales. An AGN origin of fast outflows is also favoured by the fact that feedback models of pure star-formation do not reproduce velocities larger than 500-600 km/s \citep{murray2005, lagos2013, brusa2015a, genzel2014}.

AGN-driven outflows are thought to be multiphase multiscale phenomena. 
On nuclear scales, mildly ionized warm absorbers are observed in X-rays with velocities of $\sim$500-1000 km s$^{-1}$ \citep[e.g.][]{piconcelli2005}, ultra-fast outflows (UFOs) at velocities up to 0.2-0.4c are observed through highly ionized Fe K-shell absorption \citep[e.g.][]{tombesi2011, tombesi2015}, ionized outflowing gas is observed through optical/UV Broad Absorption Line (BAL) systems in QSOs \citep[e.g.][]{dai2008}. 
On galactic scales, broad [OIII]$\lambda$5007 and H$\alpha$, that cannot be explained solely by gravitationally bounded internal motions, have been usually interpreted as evidence of galactic-scale winds up to velocities of a few thousands km s$^{-1}$ \citep[e.g.][]{forsterschreiber2009, forsterschreiber2014, harrison2012, zakamska2014, feruglio2015, perna2015a, cresci2015, brusa2015a, brusa2016, cicone2016}. 
Shifted broad components with similar velocities have been also detected in molecular lines \citep[e.g.][]{feruglio2010, sturm2011, cicone2014}. 
Finally, evidence of neutral/low-ionization outflowing gas is provided by the observation of blue-shifted UV inter-stellar medium (ISM) absorption lines \citep{hainline2011, cimatti2013}. 

Molecular and ionised outflows are characterised by mass-loss rates which can exceed the star formation rate (SFR) by two orders of magnitude and by gas depletion timescales much shorter than typical galaxy timescales \citep{cicone2014}. Also, \citet{cimatti2013} suggest a causal relation between the presence of fast galactic winds in AGN host galaxies at z$\sim$2.2 and the concomitant migration of galaxies onto the \emph{red sequence} and the decrease in the activity of moderately luminous AGN. These results suggest that AGN-driven outflows might provide the feedback invoked by the models to quench the SF. However, only few studies of high-redshift quasars have been able to provide direct evidence of feedback from the anti-correlation of the spatial distribution of the ionized phase of the outflow and star formation powered emission \citep{canodiaz2012, cresci2015, carniani2016}. 
The majority of recent studies are concentrating on the understanding of the interplay between the different phases of AGN-driven outflows and on the connection between the different scales, especially in luminous quasars. However, there is mounting evidence that also moderately-luminous AGN might be able to drive galactic-scale outflows and play a key role in feedback processes \citep{cimatti2013}.

One limitation of existing studies is that it is often difficult to place these observations into the context of the overall AGN and galaxy populations as the samples are small and inhomogeneous. Systematic studies of consistently selected samples of normal SFGs and AGN are still quite few both at low \citep[e.g.][]{concas2017} and high redshift \citep[e.g.][]{hainline2011, cimatti2013}. 
In our work we concentrate on the study the neutral/low-ionization phase of the galactic-scale outflow in a sample of high-redshift ($1.7{<}z{<}4.6$) galaxies hosting moderately luminous ($L_{X}{<}10^{45}$erg s$^{-1}$) AGN, and in a control sample of normal SFGs with no sign of AGN activity, in order to study the possible influence of the AGN on the outflowing material. We will exploit a large compilation of optical spectra and deep X-ray observations to compare outflow properties between the two groups of galaxies.

The paper is organized as follows: in Sec. \ref{sec:dataset} we present the parent multiwavelength dataset, while in Sec. \ref{sec:sel} we describe the selection of the sample analyzed in the paper and its properties; in Sec. \ref{sec:outflows} we show the analysis of the stacked spectra of the AGN sample compared to the control sample of non-AGN SFGs, while in Sec. \ref{sec:discussion} we discuss our results. A brief summary (Sec. \ref{sec:bins}) ends the paper.

Throughout this paper, we adopt a Cosmology with $H_{0}{=}70$ km s$^{-1}$/Mpc, $\Omega_{m}{=}0.3$, $\Omega_{\Lambda}{=}0.7$ and give magnitudes in AB photometric system.

\section{The multi-wavelength dataset}\label{sec:dataset}
AGN can be identified in a variety of ways taking advantage of tracers of nuclear activity at different wavelengths \citep[e.g.][]{baldwin1981, stern2005, smolcic2008, donley2012, delvecchio2014}. In this work we focused on AGN identified through X-ray surveys, that have been proved to be the most efficient way to compile nearly unbiased samples of Compton-thin AGN \citep{brandt2005}. This selection might miss Compton-thick (CT) AGN that do not constitute, however, a dominant fraction of the AGN population \citep[e.g.][]{comastri2008, akylas2012, alexander2013}.

We selected our sample from two fields, COSMOS and GOODS-S, focusing on the areas surveyed by the Chandra X-ray observatory\footnote[1]{In the COSMOS field we searched for counterparts to X-ray sources in the {\em COSMOS-Legacy} area. Howerver, we have restricted our control sample of not X-ray sources only to the central area (i.e. C-COSMOS \citep{civano2012}) in order to maximize the spectroscopic coverage. Also, the public version of the CSTACK tool (see sec. \ref{sec:spec_class}) that we used to produce stacked X-ray maps at the positions of SFGs in the control sample includes only C-COSMOS maps.}: {\em COSMOS-Legacy} \citep{elvis2009, civano2012, civano2016} and CDF-S \citep{xue2011, luo2017}.
Our parent sample is drawn from a K-selected catalogue in order not to be biased against galaxies with low SFR, since this selection is more sensitive to mass. In particular we used the UltraVista DR1 catalogue \citep{mccracken2012, ilbert2013} for the COSMOS field, and the MUSIC catalogue \citep{grazian2006, santini2009} in the GOODS-S. At $K{<}23.8$, that is the magnitude at 90$\%$ completeness limit in both fields, our selection is sensitive to stellar masses down to log$(M/M_{\odot}){\sim}10.2$ at $z{\sim}4.0$ \citep{pozzetti2010, ilbert2013}. 

\subsection{X-ray data}\label{sec:}
The K-selected photometric catalogues were complemented with X-ray data from the catalogues of Chandra X-ray sources counterparts of \citet{luo2017} and \citet{marchesi2016} based respectively on the Chandra 7 Ms exposure in the CDF-S and on the {\em COSMOS-Legacy} \citet{civano2016} survey in the COSMOS field, the latter with an effective exposure of $\sim$160 ks. The association of counterparts to the X-ray sources is detailed in the cited papers. De-absorbed rest frame 2-10 keV luminosities ($L_{X}$ hereafter) where directly taken from the {\em COSMOS-Legacy} catalogue, while in the CDF-S they were derived from rest-frame 0.5-7 keV luminosities tabulated in \citet{luo2017} assuming a $\Gamma{=}1.8$ power-law. 
The CDF-S exposure is complete down to log$L_{(0.5{-}8keV)}$${\sim}41.7$ erg s$^{-1}$ and log$L_{(0.5{-}8keV)}$${\sim}43.8$ erg s$^{-1}$ respectively at z$\sim$1.7 and z$\sim$4.6 \citep{luo2017}, which translates into log$L_{X}{\sim}41.5$ erg s$^{-1}$ and log$L_{X}{\sim}43.6$ erg s$^{-1}$ for Compton-thin AGN ($N_{H}{<}10^{23} cm^{−2}$). In the COSMOS field the completeness limit is log$L_{X}{\sim}43.0$ erg s$^{-1}$ and log$L_{X}{\sim}44.2$ erg s$^{-1}$ respectively at $z{\sim}1.7$ and $z{\sim}4.6$ \citep{marchesi2016b}.

By selecting our sample from two different fields we can exploit the specific strength of both Chandra surveys: large statistics in the COSMOS field and X-ray depth in the GOODS-S. The obvious drawback of this choice is that in the control sample of SFGs we cannot exclude the presence of X-ray AGN, in the COSMOS field, below the detection limit of the {\em COSMOS-Legacy} survey. We deal with this issue in the following sections (Sec. \ref{sec:xray_prop} and \ref{sec:outflows})

\subsection{Spectroscopic data}\label{sec:spec_data}
The main ingredient of our analysis is a large collection of optical spectra that, at our chosen redshift, sample the rest-frame UV range ($\lambda{\sim}$1000-2000~\AA). 
In order to study the outflow phenomenon in UV rest-frame spectra two sets of lines are needed: absorption lines of elements at different ionization stages, that are due to absorption by the ISM of the radiation coming from stars and that are the actual tracers of the supposedly moving gas, and spectral lines that define the systemic redshift of the galaxies, respect to which the outflow velocity can be measured. In the latter set of lines we include both stellar photospheric absorption lines, that obviously mark the rest frame of the stars in the galaxy, and nebular emission lines, that originate in nebular regions photoionized by radiation from massive O and B stars. 

We searched for spectroscopic counterparts of the sources in our K-selected photometric catalogues, within an angular separation of 0.5", in the following spectroscopic surveys conducted at various telescopes: ESO/FORS2 \citep{vanzella2008}, VVDS \citep{lefevre2005, lefevre2013}, \citet{szokoly2004}, ESO/VIMOS \citep{popesso2008, balestra2010}, K20 \citep{mignoli2005}, \citet{silverman2010}, GMASS \citep{kurk2013} in the GOODS-S field, IMACS-Magellan \citep{trump2009}, zCOSMOS-Bright and zCOSMOS-Deep \citep{lilly2007} in the COSMOS field, and the recently completed VIMOS Ultra Deep Survey (VUDS)\footnote[2]{$http{:}//cesam.lam.fr/vuds/DR1$} \citep{lefevre2015} in both fields. 

For each survey a confidence flag is defined that expresses the reliability of the spectroscopic redshift determination giving the range of probability for a redshift to be right. We adopted the flags scheme described in \citet{lefevre2015}, that is the native reliability flag scheme for the VVDS, zCOSMOS, and VUDS surveys, and homogenized the confidence flags of the other surveys to that scheme \citep[see also][]{balestra2010}. 
We also degraded the medium-resolution spectra (FORS2 and IMACS $R{\sim}660$;  VIMOS MR $R{\sim}580$) to match the resolution of the VIMOS low-resolution grism ($R{\sim}230$). We point out that the large majority (${\sim}94\%$) of the spectra used in our analysis actually comes from low-resolution VIMOS surveys, especially zCOSMOS-Deep and VUDS. 

\section{Spectroscopic sample selection and properties}\label{sec:sel}
The high S/N required to robustly measure stellar photospheric and ISM absorption lines is not achieved by currently available optical spectra. Therefore we decided to use spectral stacking techniques to study the average outflow properties of our sample with respect to AGN activity.

First, we selected spectra with a secure redshift, i.e. a high confidence flag (3, 4) corresponding to a probability ${>}95\%$ of the redshift to be correct \citep{lefevre2015}
We refined the selection with a redshift cut $1.7{<}z{\leq}4.6$ in order to sample the rest-frame range $\lambda{\sim}$1000-2000~\AA~, where the strong UV ISM lines of our interest are located. Finally, we requested that each spectrum covers at least 60$\%$ of our chosen rest-frame wavelength range in order to be included in the sample.
After this pre-selection we ended up with 1907 galaxies, 267 of which are X-ray detected.

\subsection{Spectroscopic classification}\label{sec:spec_class}

We excluded from further analysis 170 spectra classified as Type 1 AGN (TY1), i.e. showing broad (FWHM$>1900$~km s$^{-1}$) high-ionization emission lines, e.g. CIV$\lambda$1550 \citep{szokoly2004, brusa2010, hainline2011}, 161 of which X-ray detected. TY1 were excluded because in their spectra, both individual and composites, ISM absorption lines were not detectable, probably because of the strength of the AGN contribution to the UV continuum\footnote[3]{We checked this statement by actually creating the composite spectrum of all TY1 in our sample, adding also 28 TY1 from the SDSS in the same fields \citep{alam2015}, and we verified that no ISM absorption line could be detected.}.
We also excluded from further analysis 13 X-ray sources whose spectra show prominent absorption features in highly ionised lines (BAL QSOs), since these absorptions trace winds at sub-parsec scales close to the accretion disk region, therefore a different phenomenon from the galaxy-wide winds that are the object of our study \citep{murray1995, proga2004}.

In the control sample of SFGs we identified 20 objects, all located in the COSMOS field, whose spectra show narrow high-ionization UV lines. Emission lines ratios are on average consistent with the presence of an AGN \citep[][and ref. therein]{allen1998, feltre2016}. X-ray stacking of these sources, based on the CSTACK tool\footnote[4]{$http{:}//cstack.ucsd.edu/cstack$. The public version of the CSTACK tool only includes C-COSMOS and Chandra 4Ms maps, respectively in the COSMOS and CDFS fields.}, shows a ${\sim}3\sigma$ detection in the soft band. The detected flux implies an average $L_{X}{\sim}10^{42.9}$ erg s$^{-1}$ at mean redshift $z{\sim}2.4$, which is above the threshold used in \citet{xue2011} to identify luminous AGN \citep[see also][]{ranalli2003, bauer2004}. From these evidences we conclude that these galaxies are likely hosting an AGN. 

After these additional selection steps our sample was comprised of 1724 galaxies divided into the following sub-sample: 1611 SFGs with no evidence of AGN activity neither from X-ray nor from emission lines, 93 X-ray detected sources and 20 sources with no X-ray individual detection but whose mean spectral and X-ray properties are consistent with the presence of a TY2 AGN. 

\subsection{Redshift refinement and final selection}\label{sec:zref}
In our sample the redshifts had been originally measured by different people using various techniques, therefore we decided to generate a more homogeneous set of redshifts before stacking. Systemic features (stellar and/or nebular lines) can not be detected in most of the spectra, therefore we adopted a recursive procedure based on the cross-correlation \citep{tonry1979} of each single spectrum with a template, as follows. A stacked spectrum is created from the entire sample as described in Sec. \ref{sec:outflows} and it is used as template to run the cross-correlation procedure. After the first run the redshifts of the single spectra are updated and a new template is created. The process is replicated until reaching convergence. 
We excluded the Ly$\alpha$ line from the wavelength range used in the cross-correlation because the wide range of width and profiles that the line can display could bias the procedure, and masked other strong emission lines. 

For ${\sim}17\%$ of the sample, almost all Ly$\alpha$ emitters, the cross-correlation procedure failed to recover a redshift due to low S/N in the continuum combined with the intrinsic faintness of ISM lines typical of strong Ly$\alpha$-emitters \citep{shapley2003}: these galaxies were excluded from the sample. We attribute the cross-correlation failure to the fact that it is more likely to give a high-confidence flag to the redshift of a low-S/N spectrum if it is a Ly$\alpha$-emitter than otherwise \citep{cassata2015}. 

At the end of our selection the final sample used in our analysis counts 1429 galaxies, 1322 of which show no evidence of AGN activity neither from X-ray nor from emission lines, 87 are X-ray detected sources and 20 showing narrow high-ionization UV emission lines do not have X-ray individual detection but their spectral and mean X-ray properties are consistent with the presence of a TY2 AGN.
	\begin{figure}
	\centering
	\includegraphics[scale=0.45]{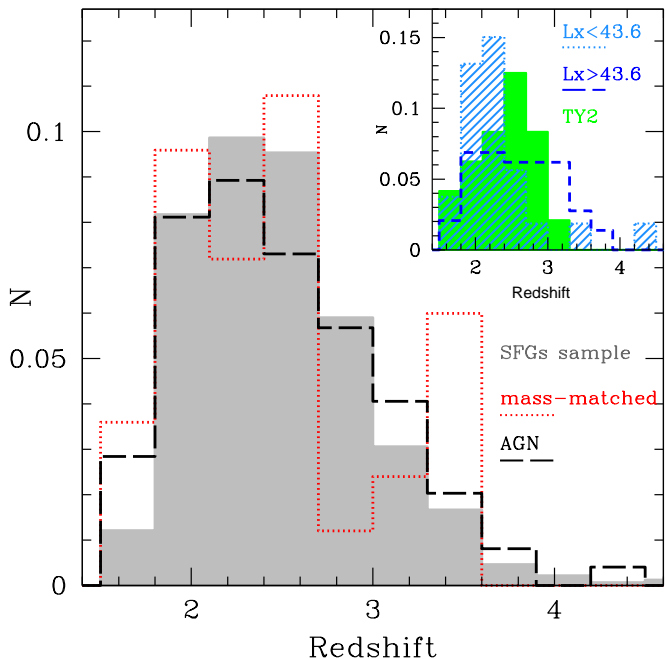}
	\caption{Normalised redsfhit distribution of the sample analysed in this work (1429 galaxies). Main panel: the grey histogram represents the control sample of SFGs (see sec. \ref{sec:xray_prop}); the black (dashed) histogram represents the AGN sample (see sec. \ref{sec:xray_prop}); the red (dotted) histogram represents the mass-matched sample of SFGs (see sec. \ref{sec:phys_prop}). Inset: redshift distribution of the AGN sub-samples. The green (shaded) histogram represents the TY2 AGN with no X-ray individual detection; the light blue (dotted) histogram represents the low-Lx sub-sample; the blue (dashed) histogram represents the high-Lx sub-sample.}
	\label{z_hist}
	\end{figure}

\subsection{X-ray properties and final classification}\label{sec:xray_prop}
Among the 87 sources in the X-ray sub-sample, 45 show the presence of narrow (FWHM${<}1900$~km s$^{-1}$) high-ionization emission lines in their spectra, e.g. CIV$\lambda$1550. Their emission lines ratios, X-ray luminosity (${<}log(L_{X}){>}{=}44.0$ erg s$^{-1}$) and hardness ratio (${<}HR{>}{\sim}0.2$, HR being defined as H-S/H+S where H and S are the count rates in the hard and soft bands, respectively) clearly indicate the presence of a Type 2 AGN (TY2).
	\begin{figure}
	\centering
	\includegraphics[scale=0.45]{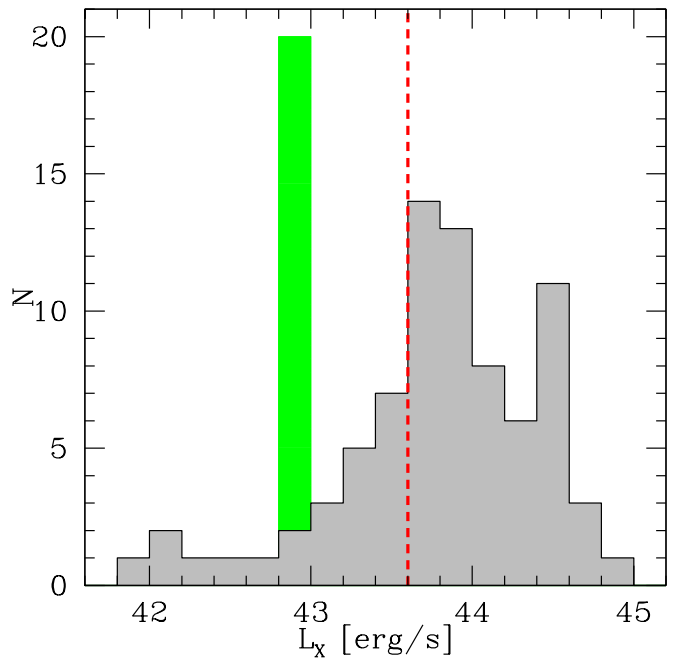}
	\caption{Distribution of the X-ray luminosities (in logarithmic scale) of the 79 X-ray detected galaxies in our sample. The red dashed line marks the separation between the two $L_{X}$ bins discussed in Sec. \ref{sec:bins} (log$L_{X}{=}$43.6 erg s$^{-1}$). The green columns marks the mean $L_{X}$, measured on stacked images (see sec. \ref{sec:spec_class}) of the 20 TY2 AGN with no X-ray individual detection.}
	\label{lx}
	\end{figure}

On the other hand, 42 spectra in the X-ray subsample do not show high-ionization emission lines. The $L_{X}$ of these sources spans a range between $10^{42}$ and $10^{44}$ erg s$^{-1}$ and its origin could be either the AGN activity or star formation. Following the \citet{luo2017} criteria 6 of those objects, characterized by $L_{X}{<}$42.3 erg s$^{-1}$, were initially classified as "Galaxies" and the other 36 as "AGN". As an additional check of this classification we converted $L_{X}$ to SFR (following \citet{ranalli2003} and assuming a Chabrier IMF) and compared the results to the SFR estimated from SED-fitting (see sec. \ref{sec:phys_prop}). We found the SFR$_{L_{X}}$ of the 6 "Galaxies" to be similar to the SED-fitting estimate, suggesting that star formation alone is enough to produce the observed $L_{X}$. Following this criterion we also changed the classification of other two objects with $L_{X}{\sim}$42.3 erg s$^{-1}$ from "AGN" to "Galaxies". For the remaining 34 objects we found the SFR$_{L_{X}}$ to be on average $\sim$1 dex higher than the SED fitting estimate, consistently with the "AGN" classification. Moreover, these 34 galaxies have ${<}HR{>}{\sim}0$, that is too flat to be explained by star formation only and consistent with an absorbed power-law with $\Gamma{=}1.8$ and column density N$_{H}{\sim}10^{22}$ cm$^{-2}$, which is typical of an obscured AGN \citep{daddi2007, cimatti2013}. 
We include the 8 objects classified as "Galaxies" based on their X-ray properties in the control sample of SFGs, while we define the 34 AGN as \emph{"no emission lines AGN"}.

Our final sample of AGN counts 99 objects: 45 have been classified as TY2 AGN based both on their UV spectra and X-ray properties, 34 have been classified as AGN based on their X-ray properties though they do not show AGN features in their UV spectra (\emph{"no emission lines AGN"}), and 20 do not have X-ray individual detection but they show narrow emission lines in their spectra and their mean X-ray properties are consistent with the presence of a TY2 AGN (see sec. \ref{sec:spec_class}).

Our final control sample of SFGs counts 1330 objects, 8 of which detected in the X-ray. The redshift distribution is shown in Fig. \ref{z_hist}, while the $L_{X}$ distribution of the AGN sample is shown in Fig. \ref{lx}.

Finally, we checked whether the control sample of SFGs is indeed free of X-ray AGN. We had already left out from the control sample galaxies showing high-ionization emission lines that are likely to host an AGN with $L_{X}$ below the Chandra detection limit (Sec. \ref{sec:spec_class}), but \emph{"no emission lines AGN"} not individually detected in the C-COSMOS maps might still be in the control sample. We built stacked images in the soft and hard bands of C-COSMOS and did the same for CDF-S (1152 and 170 sources respectively, excluding also the 8 X-ray detected SFGs). In the CDF-S we did not find any significant detection.
In C-COSMOS we found a mildly significant (2$\sigma$) detection only in the soft band. The detected soft-band flux implies an average $L_{X}{\sim}10^{42.0}$ erg s$^{-1}$ at mean redshift $z{\sim}3$, that translates into an average SFR broadly consistent with the mean SFR obtained from SED fitting. However, this X-ray emission could also come mainly from a small fraction of low-luminosity AGN. In the CDF-S the fraction of \emph{"no emission lines AGN"} with $L_{X}$ below the C-COSMOS detection limit, with respect to the control sample of SFGs, is ${\sim}7\%$ with a mean log$(L_{X})$ of 43.8 erg s$^{-1}$. If we assume that the mean X-ray emission in the stacked C-COSMOS map were produced only by ${\sim}7\%$ of the control sample galaxies in the COSMOS field their typical $L_{X}$ would be log$(L_{X}){\sim}43.15$ erg s$^{-1}$, which is broadly consistent with the mean $L_{X}$ of \emph{"no emission lines AGN"} in the CDF-S. We conclude therefore that our control sample of SFGs is likely including a small fraction of low-luminous \emph{"no emission lines AGN"}. We point out that this will not affect our results, as discussed in Sec. \ref{sec:selbias}. 

\subsection{Physical properties of the final sample}\label{sec:phys_prop}
Rest-frame colours and physical parameters were computed through SED fitting as described in \citet{ilbert2013}\footnote[5]{Though the sample selection was done on the K-selected UltraVista DR1 catalogue, for the SED fitting we used the more recent photometry presented in the COSMOS15 catalogues \citep{laigle2015}, while in the CDF-S field we complemented the MUSIC catalogues with MUSYC photometry \citep{cardamone2010}.} using only galaxy templates. In the case of narrow-lined AGN and \emph{"no emission lines AGN"}, with only an obscured view of the central engine, we do not expect that the AGN emission strongly contributes to the overall UV continuum \citep{assef2010}, therefore an SED fitting technique that considers only the galaxy component can be used to obtain a fair estimate of the colours and physical parameters of obscured AGN hosts (e.g. \citet{bundy2008}; see also \citet{bongiorno2012} for a discussion on the uncertainties in the physical properties determination when assuming no AGN component in the SED fitting).
	\begin{figure}
	\centering
	\includegraphics[scale=0.45]{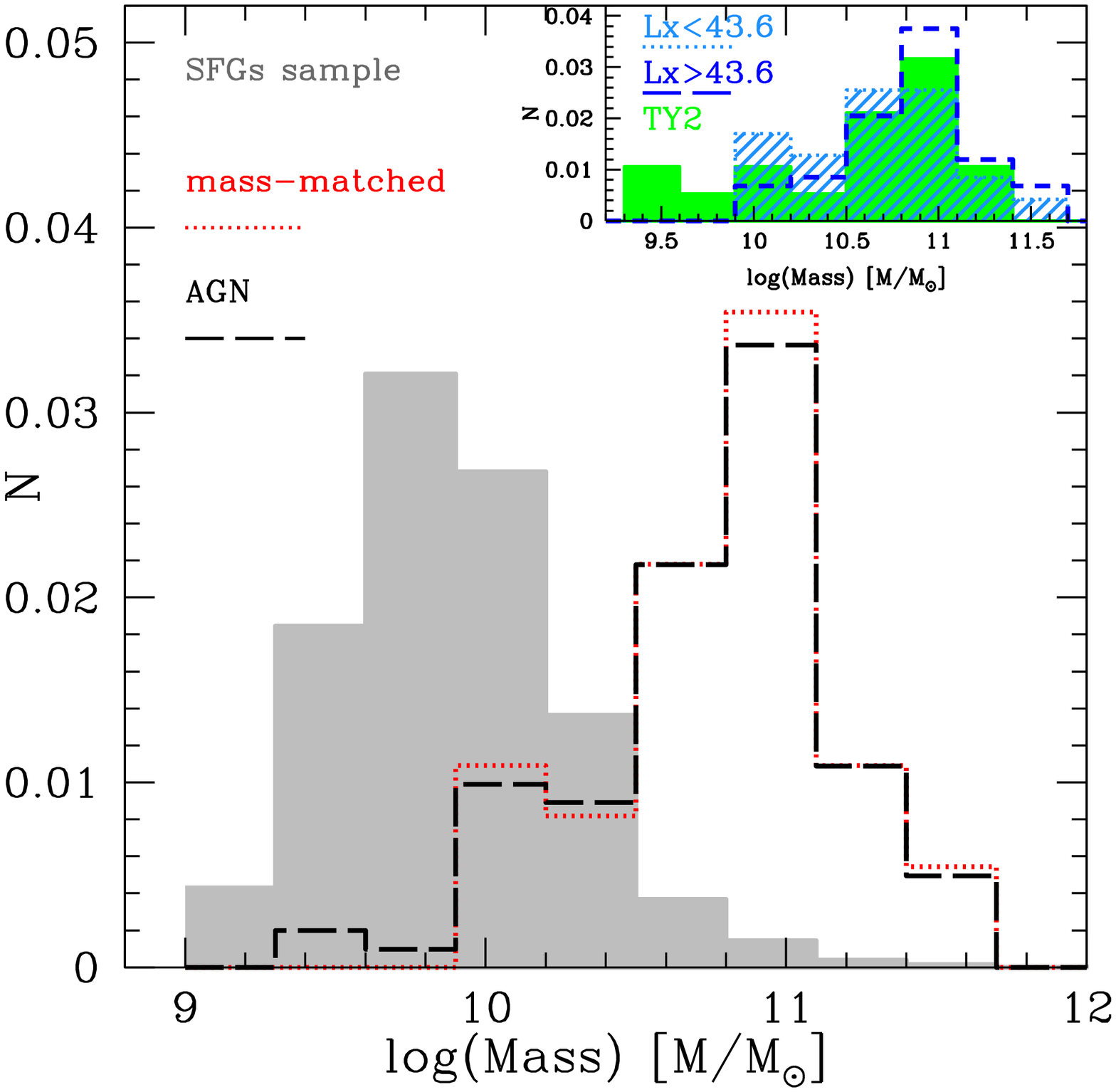}
	\includegraphics[scale=0.45]{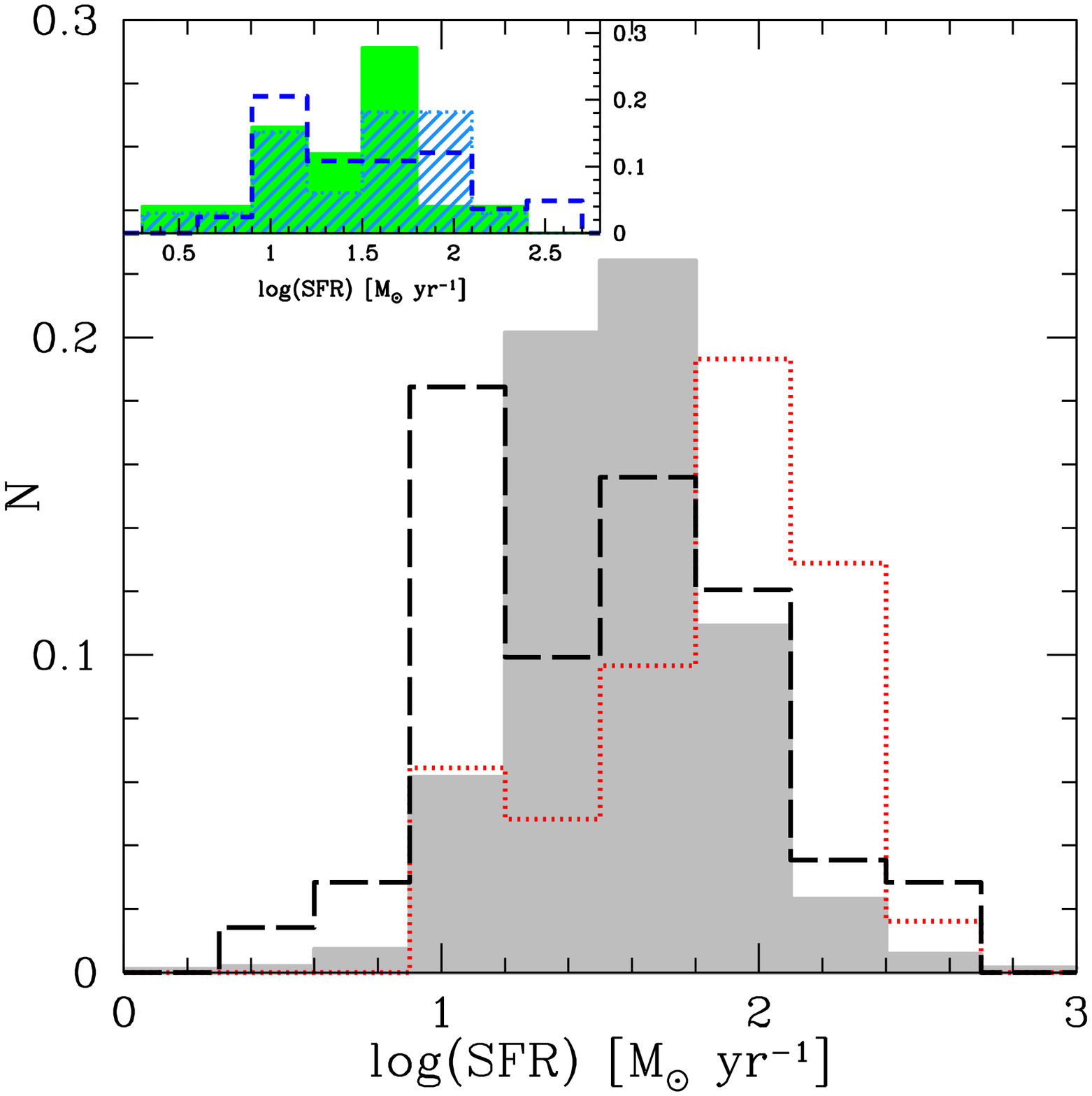}
	\caption{Stellar mass (top) and SFR (bottom) normalized distributions. Main panels: the grey histogram represents the control sample of SFGs (see sec. \ref{sec:xray_prop}); the black (dashed) histogram represents the AGN sample (see sec. \ref{sec:xray_prop}); the red (dotted) histogram represents the mass-matched sample of SFGs (see sec. \ref{sec:phys_prop}). Insets: redshift distribution of the AGN sub-samples. The green (shaded) histogram represents the TY2 AGN with no X-ray individual detection; the light blue (dotted) histogram represents the low-Lx sub-sample; the blue (dashed) histogram represents the high-Lx sub-sample.}
	\label{prop_hist}
	\end{figure}

In order to make a more fair comparison of physical properties between AGN and SFGs we also built a mass-matched control sample by selecting at random for each AGN a number of SFGs from the control sample such that the normalized distributions of stellar-mass in the two samples are the same within the uncertainties of the mass determination. The size of the mass-matched sample (34 objects with respect to a total number of 99 AGN) was limited by the nature of the stellar-mass distribution of the parent SFGs sample. 
In Fig. \ref{prop_hist} we show the distribution of stellar mass and SFR over our sample. In each plot we highlight the parameters distributions for the X-ray identified AGN samples, for the total control-sample of SFGs and for the mass-matched control sample. 
The X-ray identified AGN sample the high-mass tail of the total distribution and have on average lower SFRs than SFGs with no sign of AGN activity in the mass-matched sample\footnote[6]{A K-S test on the two SFR distributions gives a \emph{p}-value$\sim$10$^{-4}$, that implies that the null hypothesis of similar distributions is rejected at $>$99.9$\%$ significance level. }.
In Table \ref{num_gal} we summarize the mean properties of our sample in terms of redshift, mass, and SFR.
	\begin{table*}
	\centering             
	\caption{Summary of mean and median properties of the sub-samples analysed in this work. R.m.s. and M.A.D. are also shown in parenthesis. The two $L_{X}$ bins are discussed in Sec. \ref{sec:bins}.}
	\label{num_gal}
	\begin{tabular}{l l|c c c c c c c}  
	\hline
	\hline
					          && Num. & \multicolumn{2}{c}{z}     & \multicolumn{2}{c}{log(M/M$_{\odot})$} & \multicolumn{2}{c}{log(SFR) M$_{\odot}$ yr$^{-1}$}\\ 
				                  &&	 & Mean        & Median	     & Mean        & Median	 	    & Mean      & Median    \\ 
	\hline
	\multicolumn{2}{l}{SFGs}	 	  & 1330 & 2.48 (0.48) & 2.42 (0.32) & ~ 9.9 (0.4) & ~ 9.9 (0.2)             & 1.6 (0.3) & 1.6 (0.2) \\
	& \em{mass-matched SFGs}                  &   34 & 2.42 (0.53) & 2.40 (0.40) &  10.7 (0.4) & 10.8 (0.3)              & 1.8 (0.4) & 1.9 (0.2) \\
	\multicolumn{2}{l}{AGN}	                  &   99 & 2.48 (0.54) & 2.41 (0.42) &  10.7 (0.4) & 10.8 (0.2)              & 1.5 (0.5) & 1.5 (0.4) \\
	& \em{log$(L_{X}){<}43.6$ erg s$^{-1}$}   &   23 & 2.32 (0.59) & 2.21 (0.23) &  10.7 (0.4) & 10.8 (0.3)              & 1.5 (0.4) & 1.6 (0.3) \\
	& \em{log$(L_{X}){>}43.6$ erg s$^{-1}$}   &   56 & 2.58 (0.54) & 2.58 (0.43) &  10.8 (0.4) & 10.9 (0.3)              & 1.5 (0.5) & 1.5 (0.4) \\
	& \em{No X-ray TY2 AGN}      		  &   20 & 2.38 (0.36) & 2.43 (0.32) &  10.6 (0.6) & 10.7 (0.3)              & 1.4 (0.4) & 1.5 (0.2) \\
	\hline\hline
	\end{tabular}
	\end{table*}

\begin{figure}
	\centering
	\includegraphics[scale=0.45]{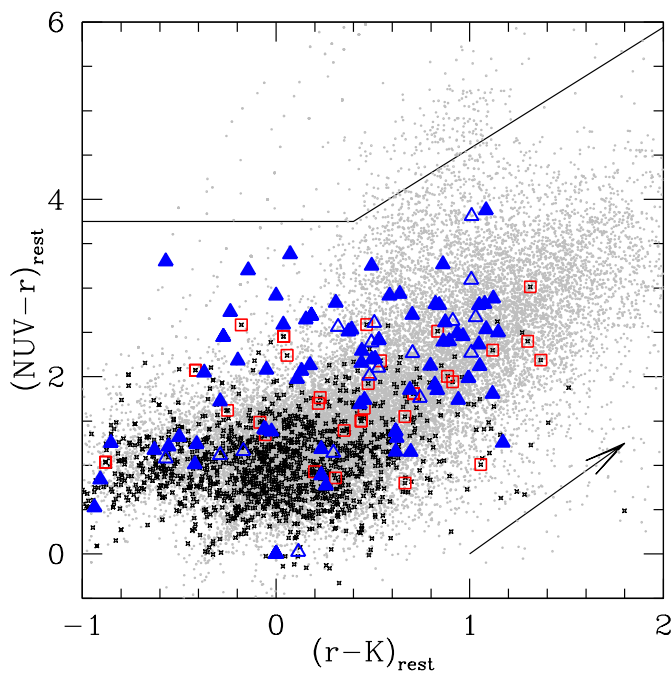}
	\caption{\emph{(NUV-r)vs.(r-K)} colour diagram. Grey dots indicate all galaxies at $1.7{<}z{<}4.6$ in the parent photometric sample ($\sim$25000 objects). Black crosses indicate the spectroscopic control sample of SFGs analysed in this work. Red squares indicate the mass-matched control sample of SFGs. Filled and empty blue triangles indicate the AGN in our spectroscopic sample, respectively AGN with an individual X-ray detection and TY2 AGN with no individual X-ray detection (identified as AGN through X-ray stacking). The heavy black lines broadly delineate the region of passive galaxies, while the black arrow shows the direction of the attenuation vector \citep{arnouts2013}.}
	\label{colcol}
	\end{figure}

We compare our selection to the parent photometric sample at the same redshift through the rest-frame \emph{(NUV-r)vs.(r-K)} colour-colour diagram \citep{arnouts2013} in Fig. \ref{colcol}. Our selection is fairly representative of the population in the bluest region of the parameters space ($(r{-}K){<}0.5$ and $(NUV{-}r){<}1.5$) in terms of median stellar mass and SFR: in the parent sample log$(M/M_{\odot})_{med}){\sim}9.7$ and log$(SFR_{med}){\sim}1.3$ M$_{\odot}$ yr$^{-1}$, similarly in our selection log$(M/M_{\odot})_{med}){\sim}9.9$ and log$(SFR_{med}){\sim}1.6$ M$_{\odot}$ yr$^{-1}$. However, with respect to the entire colour-colour space our selection is biased towards bluer, less massive, and less dusty objects, as would be expected given the requirement to have high-quality optical spectroscopy. 

\section{Searching for outflows}\label{sec:outflows}
We split our total sample into different sub-samples and build stacked spectra of each of them in order to achieve a high enough S/N to measure systemic features and ISM absorption lines.

All stacked spectra were created by shifting all the individual spectra to their rest frame using the redshifts obtained as described in Sec. \ref{sec:zref}. The spectra were then scaled by the median flux in the wavelength range 1410-1510~\AA~, that was chosen to be common to all the spectra and free from strong lines, rebinned to a dispersion of 1.3 ~\AA~~ per pixel\footnote[7]{The largest dispersion of single spectra is 5.3 ~\AA~ per pixel in the observed frame. The rest-frame dispersion of the composite was set to be 1.3 ~\AA~ $\sim$ 5.3 ~\AA~${/}(1{+}{<}z{>})$, where ${<}z{>}{=}$3.1, to match the observed pixel resolution.}, and averaged without any weights. General results would not change if we used composite spectra constructed from the median of the flux values at each wavelength instead of the average.

The uncertainty in the systemic redshifts obtained through the cross-correlation procedure (sec. \ref{sec:zref}) is expected to cause a broadening of the spectroscopic features in the stacked spectra. The redshift uncertainty in the spectra of individual galaxies is therefore propagated into the uncertainty in the measurement of the centroids of the spectral lines in the stacked spectrum, since this uncertainty is proportional to the FWHM of the line \citep{lenz1992}\footnote[8]{In order to test the validity of our results against the redshift refinement method we created the composite spectrum of the X-ray detected TY2 AGN sub-sample (see Sec. \ref{sec:xray_prop}) using the redshift given by the HeII$\lambda$1640 line for the individual galaxies. Similarly, we created the composite spectrum of a sub-sample of galaxies from the control sample of SFGs with CIII]$\lambda$1909 emission using this line to set the redshift of individual spectra following \citet{talia2012} (see Sec. \ref{sec:results} for more details about the use of these lines as tracers of the systemic frame). In both cases the results of the velocity offset measurements are indistinguishable within the errors from the ones obtained when using the cross-correlation-derived redshifts to create the stacked spectra. In this work we used the latter set of redshifts since they were derived in a homogeneous way for all the galaxies in the sample.}. 

\subsection{Measurements and uncertainties}\label{sec:}
\begin{figure}
	\centering
	\includegraphics[scale=0.45, trim=0mm 95mm 0mm 0mm, clip=true]{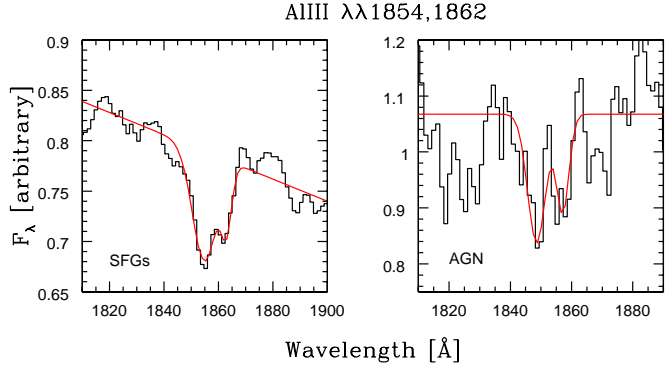}
	\caption{An example of fitted line: AlIII$\lambda\lambda$1854,1862 doublet from the stacked spectra of the control sample of SFGs (left) and of the total sample of individually X-ray detected AGN (right). In Fig. \ref{stack} the full spectra are shown. The spectral line is shown as a black histogram, while the fit (linear continuum plus gaussian profiles) is shown in red. Spectra have not been shifted to their rest-frame.}
	\label{linefit}
	\end{figure}

We measured the centroid and EW of each line on the composite spectrum of each sub-sample by simultaneously fitting each line with a Gaussian function and the local continuum with a linear function (Fig. \ref{linefit}). For each line the local continuum was defined as the wavelength range including the line and two flanking continuum bandpasses of $\sim$50~\AA~ each, corresponding to about five times the mean FWHM of the measured lines. Groups of nearby lines (closer than $\sim$50~\AA) were fitted simultaneously.

In order to estimate measurements uncertainties, for each sub-sample an artificial noiseless template was created from the quantities measured on the real composite (slope and normalization for the continuum and Gaussian parameters for the spectral lines). 
Random noise was then added to the template until the S/N of the true composite spectrum was matched, and centroids and EWs of the spectral lines of interest were measured on each realization of the noisy template. The r.m.s. of 100 repeated measurements, with respect to the input value, was adopted as the measurement uncertainty for each line. Finally, the errors in the velocity shifts were computed as the quadratic sum of the uncertainty in the measurement of the systemic lines centroid and the error in the measurement of the ISM lines centroids. 

\subsection{Determination of the systemic frame}\label{sec:systemic}
\begin{figure*}
	\centering
	\includegraphics[scale=0.67, trim=0mm 55mm 0mm 0mm, clip=true]{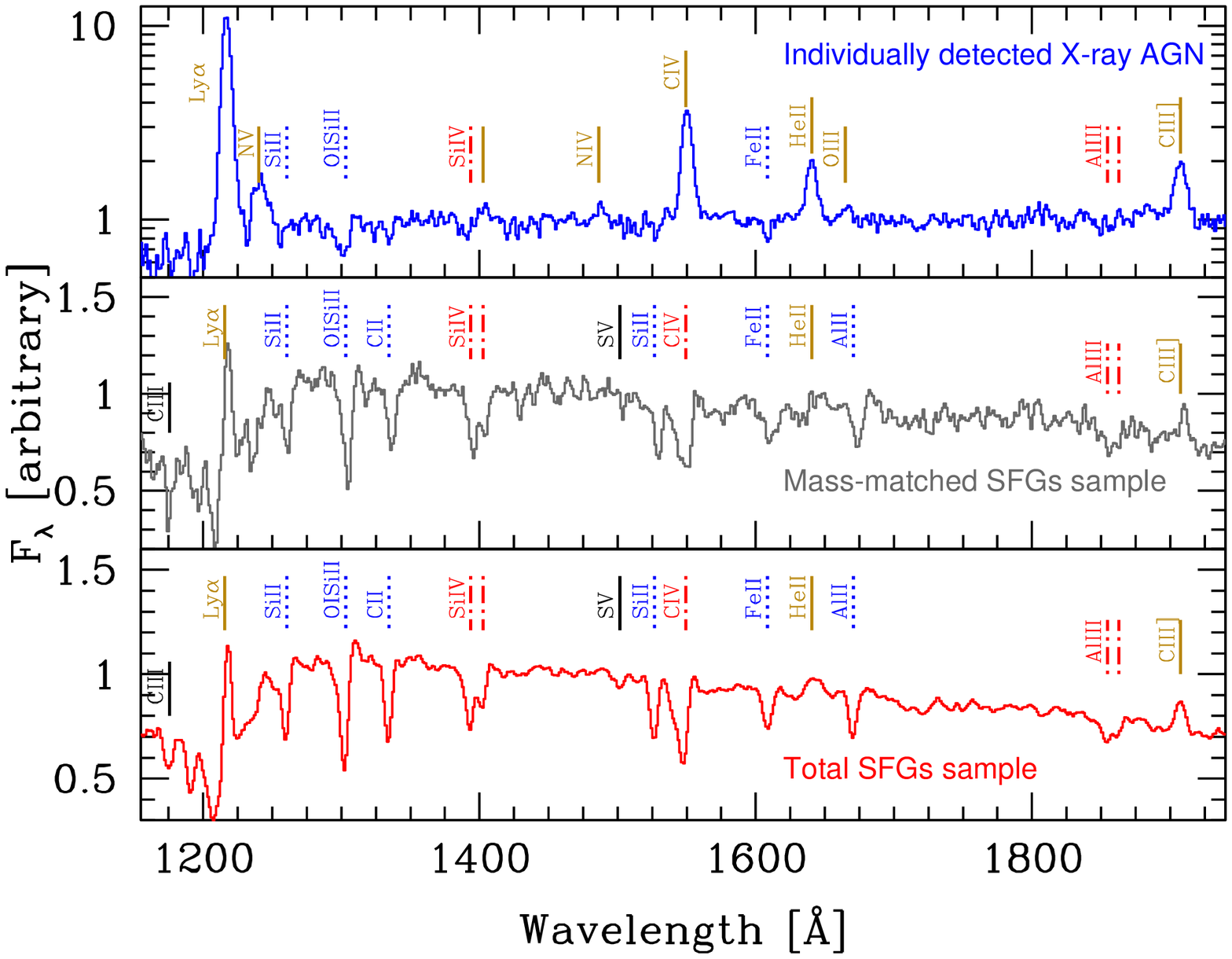}
	\includegraphics[scale=0.67, trim=0mm 55mm 0mm 0mm, clip=true]{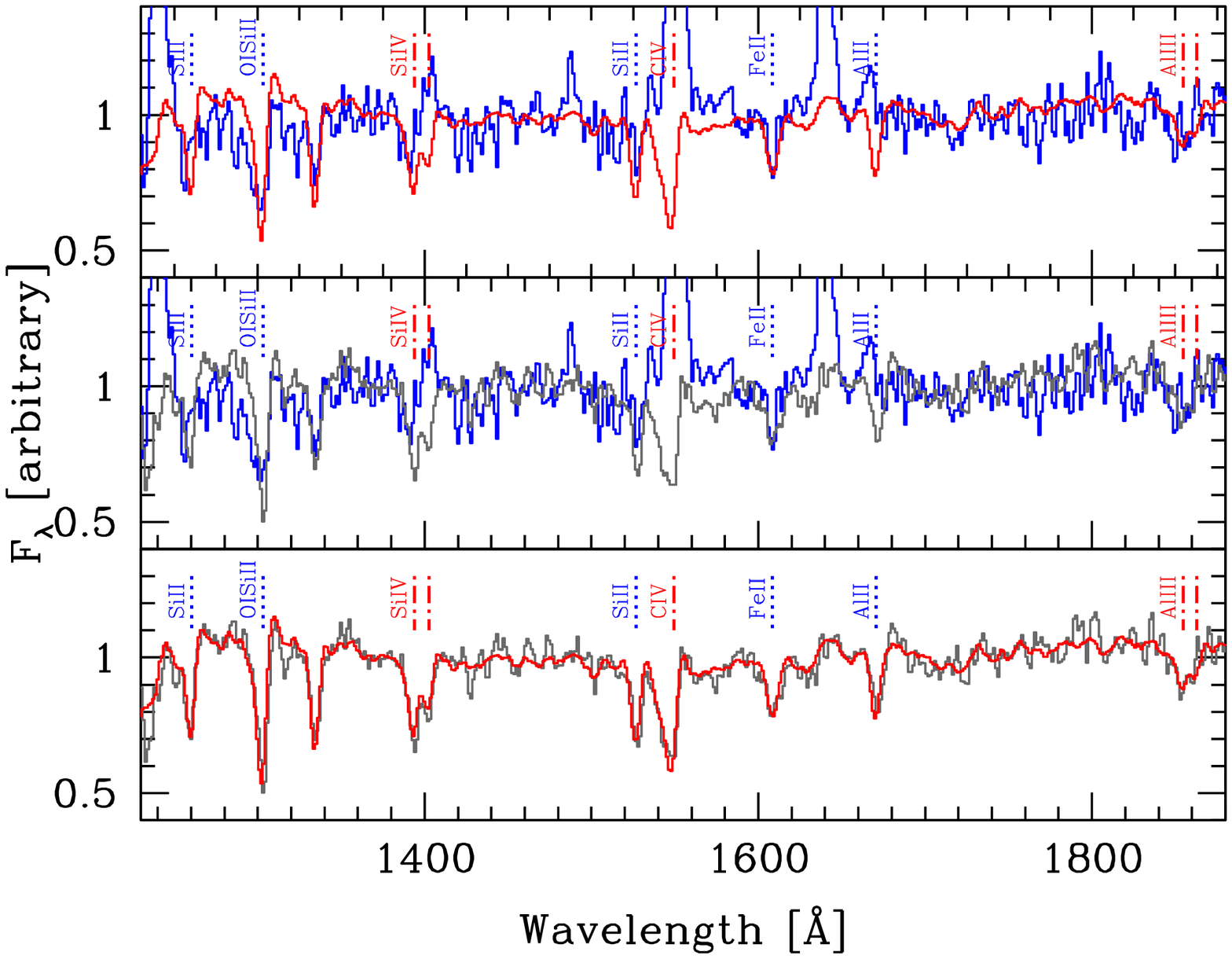}
	\caption{Top three panels, from top to bottom: stacked spectra of individually detected X-ray AGN (blue); mass-matched sample of SFGs (grey); control sample of SFGs with no sign of AGN activity (red). The AGN spectrum is shown with a flux logarithmic scale to ease visualization. 
Bottom three panels: zoom over the ISM lines to show the comparison between the continuum normalized stacked spectra of the total SFGs sample and the individually detected X-ray AGN sample (top), the mass-matched SFGs sample and the individually detected X-ray AGN sample (middle), and the total SFGs sample and the mass-matched SFGs sample (bottom). Most prominent ISM lines (except FeII$\lambda$1608) show a larger blueshift in the AGN composite spectrum with respect to both the total and mass-matched SFGs spectra.
All spectra are at rest with respect to their systemic redshift set by photospheric stellar lines (CIII$\lambda$1176 and SV$\lambda$1501) in the SFGs spectra, and the HeII$\lambda$1640 emission line in the AGN spectrum. Spectral lines of interest are marked at the position of their vacuum wavelength and labelled: photospheric stellar lines (black solid lines), emission lines (gold solid lines), ISM low-ionization absorption lines (blue dotted lines), ISM high-ionization absorption lines (red dot-dashed lines).}
	\label{stack}
	\end{figure*}

\begin{figure}
	\centering
	\includegraphics[scale=0.45]{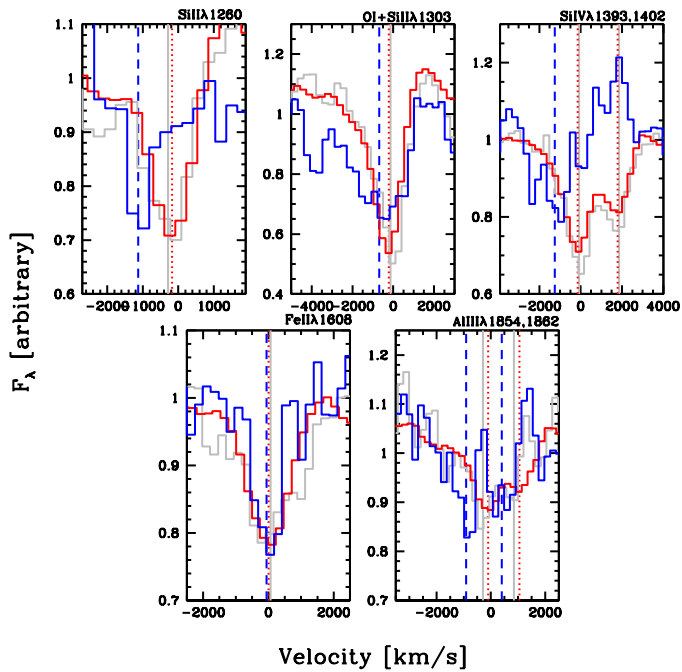}
	\caption{ISM absorption lines in the continuum normalized stacked spectra of individually detected X-ray AGN (blue); mass-matched sample of SFGs (grey); control sample of SFGs with no sign of AGN activity (red), relative to zero velocity defined by photospheric stellar lines (CIII$\lambda$1176 and SV$\lambda$1501) in the SFGs spectra, and the HeII$\lambda$1640 emission line in the AGN spectrum. For the line doublets the shortest wavelength line is taken as reference. Vertical continuous, dashed and dotted lines mark spectral lines centroids in the stacked spectra of, respectively, the mass-matched sample of SFGs, X-ray AGN, and total control sample of SFGs. Only the five lines that were measured in all three stacked spectra are shown, namely SiII$\lambda$1260, OI+SiII$\lambda$1303, SiIV$\lambda\lambda$1393,1402, FeII$\lambda$1608, AlIII$\lambda\lambda$1854,1862 (see also Table \ref{veloff}). To note that the diffent panels are on different scales on both axes.}
	\label{stack_zoom}
	\end{figure}

First of all we compare the composite spectrum of the control sample of 1330 SFGs with no evidence of AGN activity with the composite of the 79 X-ray individually confirmed AGN. 
Both stacked spectra are shown in Fig. \ref{stack}, while Fig. \ref{stack_zoom} shows more in detail the ISM lines of interest.  

Strong low-ionization absorption lines (e.g. SiII$\lambda$1260, CII$\lambda$1334, SiII$\lambda$1526, FeII$\lambda$1608, AlII$\lambda$1670) and high-ionization doublets (SiIV$\lambda\lambda$1393,1402, CIV$\lambda\lambda$1548,1550, AlIII$\lambda\lambda$1854,1862) are clearly visible in the SFGs spectrum (for a complete list of features see \citet{talia2012}). We are also able to detect stellar photospheric absorption lines (CIII$\lambda$1176 and SV$\lambda$1501) that we used to set the systemic redshift of the spectrum. 

In the AGN stacked spectrum we detect weaker low- and high-ionization absorption lines, with respect to the SFGs sample, and no photospheric stellar lines. In this case we used instead the HeII$\lambda$1640 emission line as tracer of the systemic frame. As HeII$\lambda$1640 is not a resonance line (as, for example, Ly$\alpha$ is) it can serve as a tracer for the redshift of the H II regions near/around the stars\footnote[9]{Another emission line that could be in principle be used as a tracer for the systemic redshift is the CIII]$\lambda$1909 doublet. However, the doublet is not resolved at the resolution of our sample and the flux ratio of the two components is sensitive to the electron density \citep{keenan1992}. In absence of an AGN the values that the lines ratio can take is quite narrow, therefore even in case of low-resolution spectra the doublet can be robustly modelled and used as systemic redshift tracer \citep[e.g.][]{talia2012, guaita2017}. In our SFG spectra the systemic redshift computed from the CIII]$\lambda$1909 doublet following \citet{talia2012} is in agreement, within the errors, with the one obtained from the photospheric absorption lines.
On the other hand, in presence of an AGN the knowledge of the electron density is crucial to model the emission lines doublet and this prevented us from using also this line in our analysis of the AGN composite.}.
The HeII$\lambda$1640 line can be originated by different mechanisms, e.g. PopIII stars, AGN activity, stellar winds from Wolf-Rayet (W-R) stars \citep{cassata2013}. In our X-ray AGN composite the strength of the line with respect to the SFGs composite ($\sim$10$\times$ higher) and the relative strength of the HeII$\lambda$1640 with respect to the other emission lines in the spectrum strongly favours an AGN origin over the other two possibilities, though with our data we cannot completely rule out a possible contribution from W-R stellar winds. 
The HeII$\lambda$1640 line was already successfully used to determine the systemic frame of Type 2 AGN by \citet{hainline2011} who demonstrated that the average velocity difference between HeII$\lambda$1640 and H$\alpha$ is -37 km s$^{-1}$, that is lower than our velocity uncertainties. 

\subsection{Results}\label{sec:results}
The velocity shifts of the ISM lines in the SFGs and AGN stacked spectra are reported in Table \ref{veloff}.

In the SFGs spectrum we measure velocity offsets for the ISM lines that are on average of the order of ${\sim}-150$~km s$^{-1}$. These velocities are consistent with similar studies of SFGs at $z{>}1$ \citep[e.g.][]{shapley2003, vanzella2009, talia2012, bordoloi2013}. 
The largest velocity is measured for the CIV doublet (${\sim}-650$~km s$^{-1}$), which is also the most uncertain given the likely dominating contribution of stellar winds to the line profile \citep{walborn1984}. Moreover, there is an additional uncertainty on the true centroid of the blend since in our spectra the CIV doublet is not resolved.  
The OI+SiII$\lambda$1303 measurement is also affected by the uncertainty of being an unresolved blend of two lines and by possible contamination of SiIII$\lambda$1296 stellar absorption \citep{chandar2005}.
The shifts of the other high-ionization doublets (i.e. AlIII and SIV), whose profiles are likely dominated by the interstellar component instead of stellar winds \citep{walborn1984}, are instead in line with the ones measured for the low-ionization lines. 
The FeII$\lambda$1608 is the only ISM line at rest with respect to the systemic redshift \citep[see also][]{shapley2003, talia2012}. In \citet{talia2012} it was argued that a blend of absorption features (FeIII, AlIII, NII) typical of O and B stars \citep[][]{kinney1993} could be contaminating the FeII line profile. Higher resolution spectra would be needed to understand whether the FeII ion has the same kinematics of the other low-ionization lines or if it is truly not participating in the moving gas flow.

In the AGN composite spectrum we measure velocity offsets on average of the order of $\sim-950$~km s$^{-1}$ that are much higher than in the SFGs spectrum (see Fig. \ref{stack_zoom}). We have to point out that the offsets that we measure from the ISM lines have to be considered only as lower limits, because the blueshifts account only for the radial velocity component of the outflow and with our data we cannot investigate the effects of different orientations of the outflow on the lines profile. Interestingly the FeII$\lambda$1608 line is at rest also in this spectrum. We interpret the difference in the velocity offsets of ISM lines in the two spectra as evidence that in AGN-hosts the higher outflow velocity is the result of the combined contributions of star-formation and AGN activity \citep[see also][]{hainline2011, cimatti2013}. This result is in line with simulations that indicate that pure star-formation cannot produce outflows faster than $\sim600$~km s$^{-1}$ \citep{thacker2006}. 

	\begin{table*}
	\centering             
	\caption{Velocity offsets and rest-frame EWs of ISM absorption lines in the stacked spectra. The values for the SiIV$\lambda \lambda$1393,1402 and the AlIII$\lambda\lambda$1854,1862 doublets are the average of the two components, while the values for the CIV$\lambda \lambda$1548,1550 doublet refer to the blended doublet. Significant velocity offsets ($\geq2.5\sigma$) are highlighted in boldface.} 
	\label{veloff}
	\begin{tabular}{c c r r r r r r} 
	\hline
	\hline
	Ion & Vacuum wavelength  &\multicolumn{3}{|c|}{$\Delta v$ (km s$^{-1}$)}             & \multicolumn{3}{c}{EW (\AA)}\\
	\hline
	& 			 & SFG (All)        & Mass-matched SFGs & X-ray AGN	     & SFG  (All)    & Mass-matched SFGs & X-ray AGN       \\  
	\hline
	SiII & 1260.42 		 & {\bf-176$\pm$33} & -286$\pm$139      & {\bf-1130$\pm$170} & 1.88$\pm$0.04 & 1.65$\pm$0.16 & 0.60$\pm$0.26 \\
	OI+SiII & 1303.27        & {\bf-218$\pm$32} & -131$\pm$115      & {\bf -694$\pm$200} & 4.08$\pm$0.10 & 3.44$\pm$0.22 & 3.21$\pm$0.49 \\
	CII & 1334.53            & {\bf-144$\pm$33} &   30$\pm$139      &                    & 2.28$\pm$0.05 & 2.60$\pm$0.20 &                     \\
	SiIV  &1393.76,1402.77   & {\bf-150$\pm$37} &  -78$\pm$138      & {\bf-1260$\pm$300} & 1.54$\pm$0.10 & 1.80$\pm$0.20 & 0.59$\pm$0.22       \\
	SiII & 1526.71           &      -78$\pm$40  &  143$\pm$133      &                    & 1.87$\pm$0.05 & 2.08$\pm$0.19 &         	         \\
	CIV   &1549.49           & {\bf-654$\pm$50} & -732$\pm$143      &                    & 3.90$\pm$0.12 & 4.25$\pm$0.31 &        	         \\
	FeII & 1608.45           &        5$\pm$40  &   65$\pm$227      &       -55$\pm$173  & 1.57$\pm$0.06 & 2.03$\pm$0.44 & 0.90$\pm$0.28         \\
	AlII & 1670.79           & {\bf-100$\pm$36} &  156$\pm$162      &                    & 1.20$\pm$0.05 & 1.22$\pm$0.17 &         	         \\
	AlIII &1854.72,1862.79   &      -96$\pm$50  & -296$\pm$188      & {\bf -905$\pm$190} & 0.90$\pm$0.12 & 0.85$\pm$0.20 & 1.21$\pm$0.20        \\
	\hline\hline
	\end{tabular}
	\end{table*}

In Table \ref{veloff} we also report the EW of the ISM absorption lines in the two stacked spectra.
The EW ratio of the two SiII transitions in the SFGs spectrum demonstrates that SiII transitions and likely all other low-ionization absorption lines are optically thick and therefore cannot be used to infer the column density of the neutral gas. However, we can make some brief considerations on the relative strength of the lines in the two stacked spectra.

In the stacked spectrum of SFGs all ISM absorption lines but AlIII$\lambda\lambda$1854,1862 are stronger than in that of the AGN. On the other hand, the SFGs spectrum shows a weaker Ly$\alpha$ emission line. An anti-correlation between the strength of Ly$\alpha$ and that of ISM lines in SFGs has been explained by \citet{shapley2003} by different gas covering fractions: galaxies with a larger covering fraction of dusty clouds suffer more extinction of the UV stellar continuum, as shown by their red continuum slope, as well as exhibit larger equivalent widths and weaker Ly$\alpha$ emission \citep[see also][]{pentericci2007, kornei2010}.
However, our non-AGN SFG composite has a bluer slope with respect to the AGN one. This discrepancy with the trends among non-AGN SFGs has been observed also by \citet{hainline2011} in their sample of TY2 AGN and might indicate that the Ly$\alpha$ flux that we observe in the AGN sample originates from both the ISM as well as the nuclear region. These two sources of Ly$\alpha$ photons may have disjoint properties with respect to the geometry of dust extinction, suppressing the trend observed among the non-AGN SFGs.

\subsection{Outflow velocity dependence on X-ray luminosity}\label{sec:bins}
\begin{figure*}
	\centering
	\includegraphics[scale=0.8, trim=0mm 55mm 0mm 0mm, clip=true]{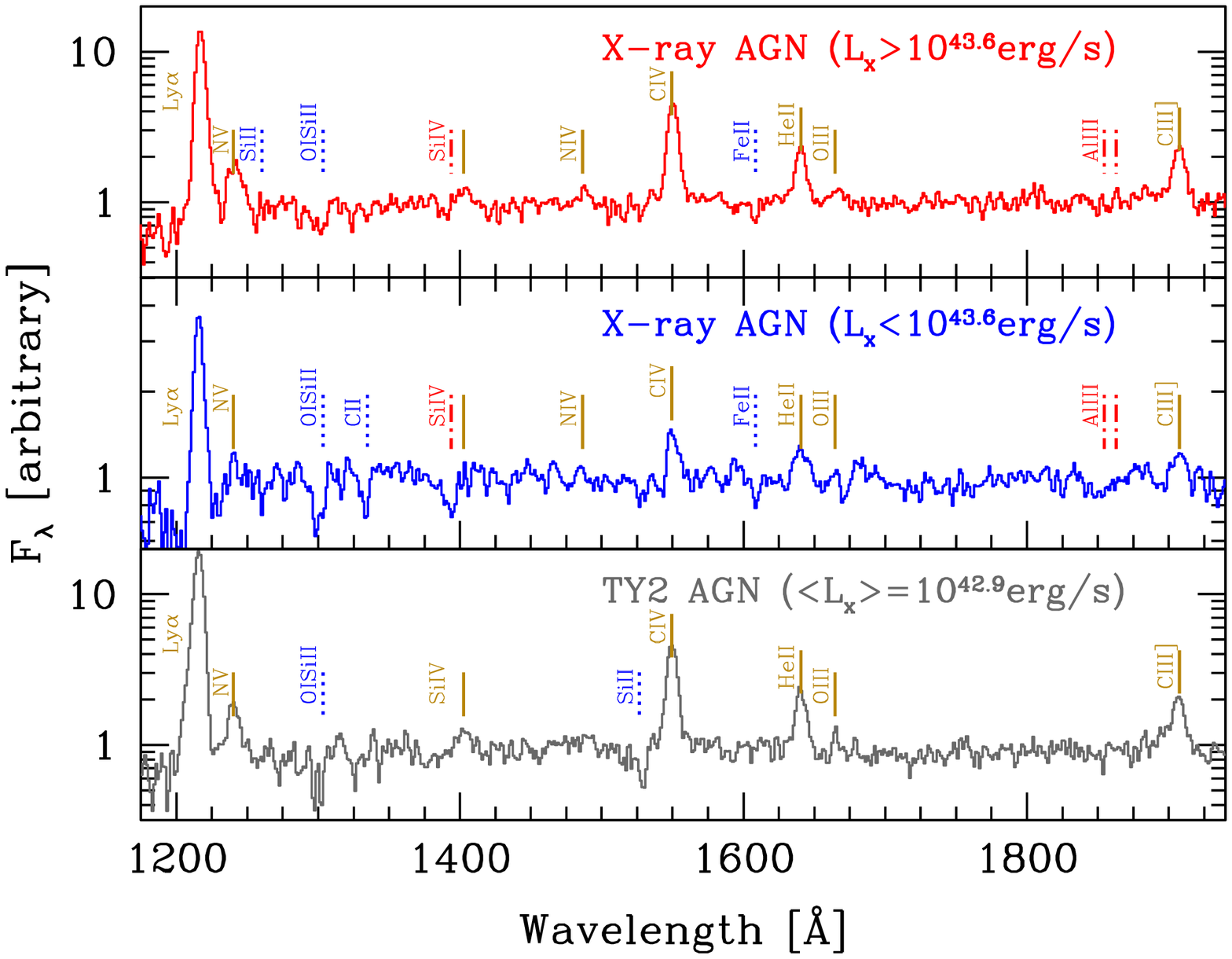}
	\includegraphics[scale=0.8, trim=0mm 125mm 0mm 5mm, clip=true]{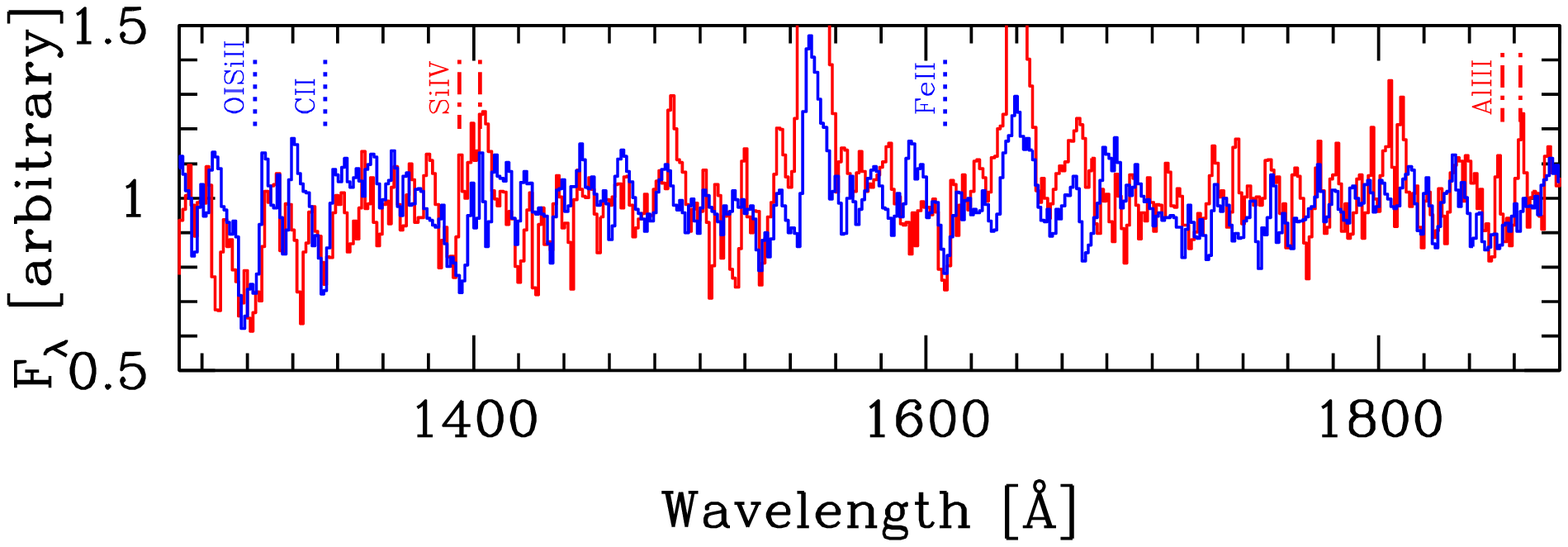}
	\caption{Top three panels, from top to bottom: stacked spectra of individually detected X-ray AGN with $L_{X}{>}10^{43.6}$ erg s$^{-1}$ (red; 56 objects); individually detected X-ray AGN with $L_{X}{<}10^{43.6}$ erg s$^{-1}$ (blue; 23 objects); TY2 AGN not individually X-ray detected (grey; 20 objects). The spectra are shown with a flux logarithmic scale to ease visualisation. Bottom panel: zoom over the ISM lines to show the comparison between the stacked spectra of two sub-samples of individually detected X-ray AGN. Most prominent ISM lines show similar blueshifts in both composite spectra.
All spectra are at rest with respect to their systemic redshift set by the HeII$\lambda$1640 emission line. Spectral lines of interest are marked at the position of their vacuum wavelength and labelled: emission lines (gold solid lines), ISM low-ionization absorption lines (blue dotted lines), ISM high-ionization absorption lines (red dot-dashed lines).}
	\label{bins}
	\end{figure*}

We further divided our X-ray AGN sample into two bins with respect to the $L_{X}$ to search for a possible dependence of the outflow velocity and built a stacked spectrum for each sub-sample as described in the previous sections. The stacked spectra are shown in Fig. \ref{bins}.
The $L_{X}$ distribution over the sample shows a peak around log$(L_{X}){\sim}43.8$ erg s$^{-1}$ and a low-luminosity tail. We decided to split the sample at log$(L_{X}){\sim}43.6$ erg s$^{-1}$: after some tests we verified that this is also the optimal binning in order to have two stacked spectra of comparable S/N in the continuum. In Table \ref{num_gal} we show that the two sub-samples have similar mean (median) stellar masses and SFRs: as discussed in the following section this ensures that a difference in velocity offsets between the two spectra should be due mainly to the different properties of the AGN in the two samples. However, as shown in Table \ref{binoff}, no significant difference is found between the two composite spectra. The largest difference is seen for the SiIV$\lambda$1393,1402 doublet, that shows a smaller EW in the high-$L_{X}$ composite spectrum, possibly due to a higher contamination from the emission line with respect to the high-$L_{X}$ bin. However the difference is not highly significant given the errors.

	\begin{figure}
	\centering
	\includegraphics[scale=0.45]{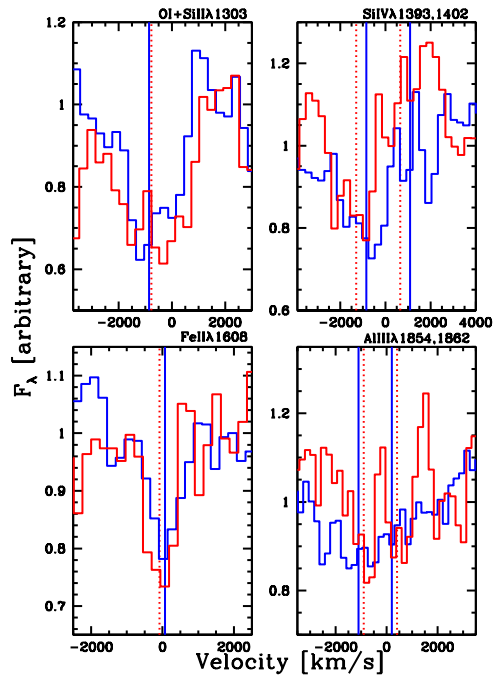}
	\caption{ISM absorption lines in the stacked spectra of individually detected X-ray AGN with $L_{X}{>}10^{43.6}$ erg s$^{-1}$ (red); individually detected X-ray AGN with $L_{X}{<}10^{43.6}$ erg s$^{-1}$ (blue), relative to zero velocity defined by the HeII$\lambda$1640 emission line. For the line doublets the shortest wavelength line is taken as reference. Vertical continuous and dotted lines mark spectral lines centroids in the stacked spectra of, respectively, the low-$L_{X}$ and high-$L_{X}$ AGN samples. Only the four lines that were measured in both stacked spectra are shown, namely OI+SiII$\lambda$1303, SiIV$\lambda\lambda$1393,1402, FeII$\lambda$1608, AlIII$\lambda\lambda$1854,1862 (see also Table \ref{binoff}). To note that the diffent panels are on different scales on both axes.}
	\label{bin_zoom}
	\end{figure}

	\begin{table*}
	\centering             
	\caption{Velocity offsets and rest-frame EWs of ISM absorption lines in the stacked spectra. The values for the SiIV$\lambda \lambda$1393,1402  and the AlIII$\lambda\lambda$1854,1862 doublets are the average of the two components.} 
	\label{binoff}
	\begin{tabular}{c c r r r r}  
	\hline
	\hline
	&  &$L_{X}{<}10^{43.6}$ erg s$^{-1}$ & $L_{X}{>}10^{43.6}$ erg s$^{-1}$ & $L_{X}{<}10^{43.6}$ erg s$^{-1}$ & $L_{X}{>}10^{43.6}$ erg s$^{-1}$\\  
	\hline
	Ion & Vacuum wavelength                 &\multicolumn{2}{|c|}{$\Delta v$ (km s$^{-1}$)} & \multicolumn{2}{c}{EW (\AA)}\\
	\hline
	OI+SiII & 1303.27                       & -850$\pm$140 & -770$\pm$250 & 3.65$\pm$0.90 & 3.32$\pm$1.05           \\
	CII & 1334.53                           & -400$\pm$200 &              & 2.40$\pm$0.70 &             	        \\
	SiIV  &1393.76,1402.77                  & -850$\pm$210 &-1290$\pm$350 & 1.94$\pm$0.90 & 0.60$\pm$0.90           \\
	FeII & 1608.45                          &   70$\pm$200 &  -83$\pm$130 & 0.90$\pm$0.32 & 1.20$\pm$0.70           \\
	AlIII &1854.72,1862.79                  &-1100$\pm$440 & -900$\pm$180 & 0.95$\pm$1.09 & 1.04$\pm$0.60           \\
	\hline\hline
	\end{tabular}
	\end{table*}

We also created the stacked spectrum of the 20 TY2 AGN with no individual X-ray detection (Fig. \ref{bins}). 
As pointed out in Sec. \ref{sec:sel} we cannot secure the presence of an AGN in the individual sources; however these objects show evidence of AGN activity in their stacked spectrum and X-ray maps, with an average <logL$_{X}$>42.9 erg s$^{-1}$, that is comparable to the average L$_{X}$ of the 23 sources in the low-luminosity bin of the X-ray confirmed AGN sample\footnote[10]{We remind the reader that all 20 TY2 AGN with no X-ray detection are located in the COSMOS field where Chandra observations were shallower than in the CDFS, while most of the 23 sources in the low-luminosity bin of the X-ray confirmed AGN sample are located in the CDFS.}. In this case only two ISM lines could be detected: OI+SiII$\lambda$1303 and SiII$\lambda$1526 for which we measure, respectively, velocity offsets of -765$\pm$195 km s$^{-1}$ and 167$\pm$170 km s$^{-1}$. The velocity offset of the OI+SiII$\lambda$1303 is perfectly in line with the velocities measured in the X-ray confirmed AGN, further confirming the absence of a correlation between outflow velocity and X-ray luminosity in AGN host galaxies. Curiously, the SiII$\lambda$1526 is instead at rest with respect to the HeII$\lambda$1640. We suspect a contamination by an absorption component of CIV$\lambda$1550, but higher resolution spectra would be required to further investigate this issue.

\subsection{Assessing the influence of potential biases}\label{sec:selbias}
The AGN in our sample populate the high-mass tail of the mass distribution of the total sample, while the control sample contains a large number of lower-mass SFGs. Therefore we built a mass-matched sample as detailed in Sec. \ref{sec:phys_prop}. Previous studies find no significant correlation between stellar mass and outflow velocity \citep{talia2012, bordoloi2013, bradshaw2013, karman2014} and in fact we measured velocity offsets in the mass-matched sample spectrum that are broadly consistent, within the errors, to the ones measured in the composite of the total SFGs sample (see Table \ref{veloff}. In particular, also for the mass-matched sample the velocity offsets are always significantly lower than in the AGN composite spectrum.

On the other hand discrepant results have been reported on the possible dependence of outflow velocity on SFR, with some authors reporting no correlation \citep[e.g.][]{talia2012} and others reporting a positive one \citep[e.g.][]{bradshaw2013}. Our AGN have a lower SFR, on average, with respect to SFGs with the same stellar mass (see Sec. \ref{sec:phys_prop} and Table \ref{num_gal}), therefore they should show similar or slower velocities with respect to the SFGs sample, if star-formation were the only outflow driving mechanism. We therefore interpret the different velocity offsets as evidence of a strong AGN contribution in powering the outflow in the galaxies of the AGN sample.
We are aware that an X-ray selection is biased against obscured sources, however this should not bias our AGN sample against galaxies with higher sSFR. For example, in \citet{lanzuisi2015} and \citet{georgantopoulos2013} CT AGN are shown to lie typically on the main sequence. Also \citet{rovilos2012} find no correlation between N$_{H}$ and sSFR. Finally, \citet{lanzuisi2017} find a correlation between N$_{H}$ and mass but not SFR. Given the arguments presented in the cited papers, we have no strong reason to believe that the missing fraction of CT AGN is affecting the SFR distribution of our sample. 

We also know that the control sample is likely to contain a small (${\sim}$7$\%$) fraction of \emph{"no emission lines AGN"} in the COSMOS field individually undetected in X-ray maps (see Sec. \ref{sec:xray_prop}). 
To evaluate how their presence could bias our results we considered only galaxies in the CDF-S. We built two stacked spectra: one only with galaxies from the control sample and the other adding also \emph{"no emission lines AGN"} in the CDF-S with $L_{X}$ below the C-COSMOS detection limit. We did not find a statistically significant difference in the velocity offsets measured in the two spectra, though with high measurements uncertainties due to the low number of spectra involved in the exercise, meaning that the percentage of galaxies with higher velocity shifts in their ISM lines is too low to significantly change the mean measurements. We do not know whether the relative number of \emph{"no emission lines AGN"} in C-COSMOS, with respect to SFGs with no AGN activity, would be the same as in CDF-S. However, we point out that even if the percentage of X-ray undetected \emph{"no emission lines AGN"} would be much higher the net effect would be to increase the velocity offsets measured in the stack of the control sample. In principle, a clean sample of "pure" SFGs would therefore show lower velocities than the ones that we have reported in Table \ref{veloff} making the outflow velocity difference between SFGs and AGN hosts even larger.

\section{Discussion}\label{sec:discussion}
We compared the velocity of the low-ionization phase of the outflow in a sample of high-redshift SFGs with and without X-ray confirmed AGN activity and found that outflows are faster in AGN host galaxies \citep[see also][]{hainline2011, cimatti2013}. Many studies of the ionized phase of the outflow report similar findings \citep[e.g.][]{harrison2012, forsterschreiber2014, perna2015a, perna2015b, brusa2015a, brusa2016}, suggesting a similar behaviour of the different phases of the outflow.

A correlation between outflow velocity and the power of the AGN would be expected since the presence of an AGN can boost the wind velocity with respect to the effect of star formation alone. This issue has been addressed by some studies of the ionized phase. For example, \citet{zakamska2014} find a trend of [OIII]$\lambda$5007 emission line width (a proxy of outflow velocity in the ionized phase) with the IR luminosity in SDSS luminous quasars, as expected for outflows driven by the radiation pressure of the quasar \citep[e.g.][]{menci2008}. \citet{perna2017} find as well a positive correlation between outflow velocity and X-ray luminosity of SDSS AGN (both TY1 and TY2). Studies like \citet{brusa2015a} that find no correlation between AGN luminosity and ionized phase wind velocity are analysing biased samples that were specifically selected to have an enhanced probability of showing fast galactic-scale winds.

\begin{figure}
	\centering
	\includegraphics[scale=0.45]{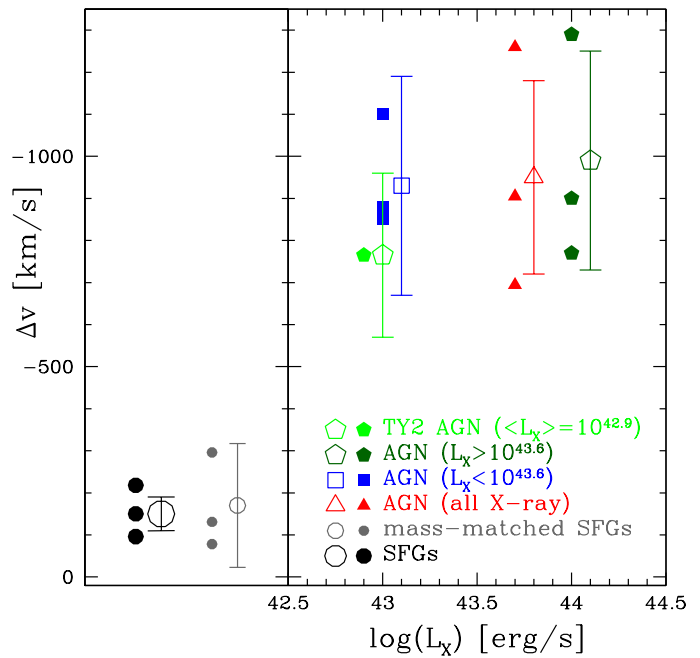}
	\caption{Measured velocity offsets of ISM lines in the six stacked spectra analysed in the paper as a function of X-ray luminosity (see Tables \ref{veloff} and \ref{binoff}): control sample of SFGs (black big circles), mass-matched sample of SFGs (grey small circles), total sample of individually X-ray detected AGN (red triangles), X-ray AGN with $L_{X}{<}10^{43.6}$ erg s$^{-1}$ (blue squares), X-ray AGN with $L_{X}{>}10^{43.6}$ erg s$^{-1}$ (dark green pentagons), TY2 AGN not individually detected in X-ray (light green pentagons). Filled symbols indicate measurements of individual lines tracing outflowing gas. We only show the blueshifted lines that are measured in all the spectra: OI+SiII$\lambda$1303, SiIV$\lambda\lambda$1393,1402, AlIII$\lambda\lambda$1854,1862 (for the TY2 AGN with no individual detection the only measured line is OI+SiII$\lambda$1303). Empty symbols indicate the mean value for each sub-sample and are slightly shifted in the x-direction to ease visualization. Error bars represent mean errors for each sub-sample.}
	\label{velplot}
	\end{figure}

Our analysis of the low-ionization phase of the ISM offers a complementary view on the outflow phenomenon in AGN hosts. In particular, our results do not show a correlation between $L_{X}$ and outflow velocity: we measure instead similar ISM absorption lines offsets in the stacked spectra of AGN with different X-ray luminosities (see Fig. \ref{bins} and \ref{velplot}).
Assuming a simple model in which UV ISM lines are produced by clouds of absorbing material that are swept by the hotter flow of more highly ionized material \citep{heckman2000}, our results would suggest that, regardless of the power of the energy source driving the outflow, these clouds can be accelerated only up to a certain velocity before being disrupted. Alternatively (or complementarily), cool clouds could be forming only at a certain distance from the energy source that is driving the wind, inside the decelerating flow. The ionized gas that produces high-ionization absorption lines would be likely located in the outer regions of the clouds producing the low-ionization lines \citep{shapley2003}. 
Multiphase studies of the outflow in local galaxies show a great variety in the relative distribution of the neutral and ionized component of the outflow suggesting that the two phases of the outflow may be mixed only up to relatively low velocities, while the highest velocities can be reached only by the highly ionized phase \citep{rupke2013}. Our results may be consistent with this picture.

In order to understand whether the measured outflow velocities would be sufficient for the gas to actually leave the galaxy, we computed the escape velocities ($V_{esc}{=}\sqrt{2GM/r_{e}}$, where \emph{M} is the sum of stellar and gas mass) expected respectively for the median mass of the non-AGN SFGs sample (log(M$_{\star}$/M$_{\odot}$)${\sim}$9.9) and of the X-ray detected AGN sample (log(M$_{\star}$/M$_{\odot}$)${\sim}$10.8). Following the prescription of \citet{magdis2012} to estimate the fraction of gas mass we derive a total gas mass log(M$_{gas}$/M$_{\odot}$)${\sim}$10.5 for the SFGs sample and log(M$_{gas}$/M$_{\odot}$)${\sim}$10.9 for the X-ray AGN sample. Assuming a typical size of $r_{e}{\sim}$5 kpc \citep{vanderwel2014} we derive $V_{esc}\sim$260 km s$^{-1}$ for the non-AGN SFGs sample and of $V_{esc}\sim$500 km s$^{-1}$ for the AGN sample. 
Comparing these values of $V_{esc}$ with the velocity offsets in Tables \ref{veloff} and \ref{binoff} it is evident that SFR-driven winds are hardly capable of reaching the outskirts of the galaxy, while AGN activity can push ISM winds beyond the limit given by the escape velocity, though in this exercise we are not considering the dark matter halo, therefore we cannot establish if the outflowing gas could be effectively injected in the inter-galactic medium \citep[see][]{harrison2012}. 
Also, whether the warm clouds embedded in the ionized outflow can actually reach the outskirts of the galaxy before disrupting or changing phase is not clear. Deep spatially resolved studies would be needed to trace the journey of the low-ionization gas inside the outflow.

However, a hint that a fraction of ISM gas in either neutral or ionized form must be escaping from the AGN hosting galaxies is given by the different average SFR that we measure in the X-ray AGN sample and in the mass-matched non-AGN SFGs sample (see also sec. \ref{sec:selbias}). We can speculate that in our sample the fact that the SFR is lower in galaxies hosting an AGN is related to their enhanced outflows that are effectively removing gas from the galaxy reducing the formation of stars. 
Up to now only few spatially resolved studies of high-redshift quasars have been able to provide evidence of feedback from the anti-correlation of the spatial distribution of the ionized phase of the outflow and star formation powered emission \citep{canodiaz2012, cresci2015, carniani2016}. 

\section{Summary}\label{sec:bins}
In this paper we have analysed a spectroscopic sample of 1429 galaxies at $1.7{<}z{<}4.6$ to study the possible relation between outflows in the low-ionized gas component and AGN activity. We have exploited Chandra data in two fields (COSMOS and CDF-S) to identify moderately luminous ($L_{X}{<}10^{45}$erg s$^{-1}$) AGN based on their $L_{X}$. We built stacked spectra in order to achieve a S/N high enough to measure robustly the velocity shifts of ISM absorption lines that are usually associated to outflowing gas.
The main results of this work can be summarized as follows.

\begin{itemize}
\item We divided our sample into X-ray confirmed AGN and SFGs with no AGN activity and measured ISM lines in the average spectra created from the two sub-samples. In the SFGs spectrum we measure ISM lines blue-shifted on average ${\sim}-150$~km s$^{-1}$, consistent with previous studies at various redshifts. This offset is interpreted as evidence of gas moving towards the line of sight. In the AGN spectrum we measure stronger velocity offsets of the order of ${\sim}-950$~km s$^{-1}$. We interpret the difference in the velocity offsets of ISM lines in the two spectra as evidence that AGN activity can boost the outflow to velocities that cannot be reached with the contributions of star-formation alone. 

\item We further divided the X-ray sample into two bins to study the possible dependence of outflow velocity on X-ray luminosity. We found no evidence of such correlation, since we measured the same average velocity offsets in the two spectra. This might indicate that the cool clouds of neutral/low-ionized absorbing gas embedded in the outflow can be accelerated only up to a maximum velocity before being disrupted. Therefore, though the AGN-powered outflows can reach velocities higher than the escape velocity of the host galaxy, it is not clear whether the neutral clouds can survive until they reach the galaxy outskirts or if all the possibly escaping gas is in ionized diffuse phase. 
\end{itemize} 

\noindent In conclusion, our data are consistent with an AGN-driven feedback picture. However, to gain further insight in the outflow phenomenon we need statistically large samples of galaxies in which both low- and high-ionization phases of the outflow can be studied simultaneously with enough S/N and spatial resolution. 

\section*{Acknowledgements}
MT whishes to thank Michele Perna for sharing a preview of his results on SDSS AGN. MT also thanks Giorgio Lanzuisi for very useful discussion on CT AGN and obscuration.  
The authors thank the anonymous referee for constructive comments.
We all thank the ESO staff for their continuous support for the VUDS survey, particularly the Paranal staff conducting the observations and Marina Rejkuba and the ESO user support group in Garching.
This work is supported by funding from the European Research Council Advanced Grant ERC--2010--AdG--268107--EARLY and by INAF Grants PRIN 2010, PRIN 2012 and PICS 2013. 
AC, OC and MT acknowledge the grant MIUR PRIN 2010--2011.  
MB acknowledges support from the FP7 grant ``eEASy'': (CIG 321913). 
RA acknowledges support from the ERC Advanced Grant 695671 'QUENCH'.
We acknowledge the grants ASI n.I/023/12/0 "Attivit\`a relative alla fase B2/C per la missione Euclid" and MIUR PRIN 2010-2011 "The dark Universe and the cosmic evolution of baryons: from current surveys to Euclid" and PRIN MIUR 2015 "Cosmology and Fundamental Physics: illuminating the Dark Universe with Euclid".
This work is based on data products made available at the CESAM data center, Laboratoire d'Astrophysique de Marseille. 
This work partly uses observations obtained with MegaPrime/MegaCam, a joint project of CFHT and CEA/DAPNIA, at the Canada-France-Hawaii Telescope (CFHT) which is operated by the National Research Council (NRC) of Canada, the Institut National des Sciences de l'Univers of the Centre National de la Recherche Scientifique (CNRS) of France, and the University of Hawaii. This work is based in part on data products produced at TERAPIX and the Canadian Astronomy Data Centre as part of the Canada--France--Hawaii Telescope Legacy Survey, a collaborative project of NRC and CNRS.



\bibliographystyle{mnras}
\bibliography{references} 


\bsp	
\label{lastpage}
\end{document}